\documentclass[prl,twocolumn,color,floatfix]{revtex4}
\usepackage{amsfonts}
\usepackage{amsmath}
\usepackage{amssymb}
\usepackage{graphicx}
\usepackage{color}

\begin{document}
\title{Energy Down Conversion between Classical Electromagnetic Fields 
via a Quantum Mechanical SQUID Ring}
\author{M.J. Everitt}
\email{m.j.everitt@sussex.ac.uk}
\author{T.D. Clark}
\email{t.d.clark@sussex.ac.uk}
\author{P.B. Stiffell}
\author{C.J. Harland}
\affiliation{Centre for Physical Electronics and Quantum Technology, 
             School of Science and Technology, 
             University of Sussex, 
             Falmer, 
             Brighton, 
             BN1 9QT, 
             U.K.}
\author{J.F. Ralph}
\affiliation{Department of Electrical and Electronic Engineering,
             Liverpool University,
             Brownlow Hill, 
             Liverpool,
             L69 3GJ, 
             U.K.}

\begin{abstract}
  We consider the interaction of  a quantum mechanical SQUID ring with
  a classical resonator  (a parallel $LC$ tank circuit).  In our model
  we  assume that  the evolution  of  the ring  maintains its  quantum
  mechanical nature, even though the circuit to which it is coupled is
  treated classically.  We show that  when the SQUID ring is driven by
  a   classical  monochromatic   microwave  source,   energy   can  be
  transferred between this  input and the tank circuit,  even when the
  frequency  ratio between  them  is very  large.  Essentially,  these
  calculations  deal with  the coupling  between a  single macroscopic
  quantum object (the SQUID  ring) and a classical circuit measurement
  device where due account  is taken of the non-perturbative behaviour
  of the ring  and the concomitant non-linear interaction  of the ring
  with this device.
\end{abstract}
\maketitle

\section{\bigskip Introduction}

With the now very serious interest being taken in the possibilities of
creating quantum  technologies such as  quantum information processing
and quantum computing~\cite{r1,r2,r3}, much attention is being focused
on the application of Josephson effect devices, particularly the SQUID
ring.  As has been  demonstrated recently,  with the  appropriate ring
circuit parameters and operating temperature these devices can display
manifestly  macroscopic  quantum  behaviour  such  as  macroscopically
distinct superposition  of states~\cite{r4,r5,r6,r7,r8,r9,r10,r11}. In
any   considerations  of   quantum  technologies   the  role   of  the
environment,  as coupled  to  the quantum  object,  has featured  very
strongly~\cite{r12,r13,r14}.  With  regard  to  using  superconducting
systems in quantum technologies, it has been shown that Josephson weak
link circuits,  and in particular  SQUID rings in the  quantum regime,
are  highly non-perturbative in  nature and  can generate  very strong
non-linear       interactions       with       classical       circuit
environments~\cite{r15}.  In this  paper we  provided  a demonstration
that this  non-perturbative (non-linear)  behaviour is crucial  to the
understanding  of   the  interaction  of  SQUID   rings  with  circuit
environments.  In this work  we first  consider the  adiabatic (ground
state) interaction of a  quantum regime SQUID ring inductively coupled
to a classical parallel resonance $LC$ (tank) circuit.
\begin{figure}[tb]
\begin{center}
\resizebox*{0.4\textwidth}{!}{\includegraphics{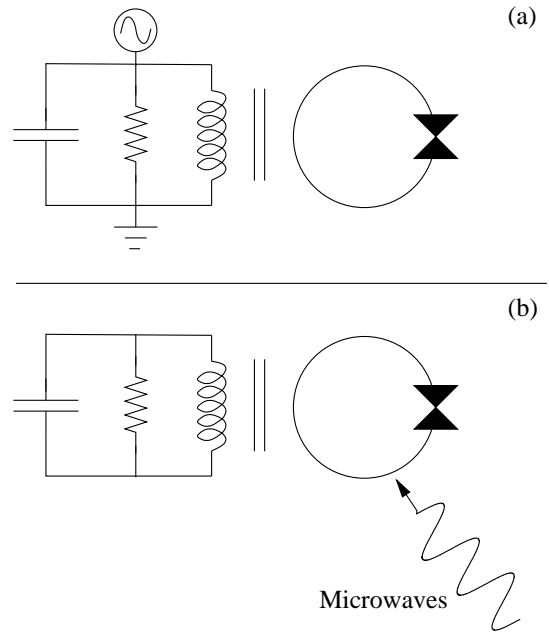}}
\caption{ 
  (a) Schematic  of a (quantum regime) SQUID  ring inductively coupled
  to a driven tank circuit (two  mode system) (b) schematic of a three
  mode  system: classical  input microwave  excitation source  - SQUID
  ring - and an inductively coupled tank circuit.}
\label{fig:schematic}
\end{center}
\end{figure}
This     circuit     system      is     shown     schematically     in
figure~\ref{fig:schematic}(a).  We then  continue  to consider  energy
transfer  between  a  classical  monochromatic microwave  source  (the
input) and this classical electromagnetic (em) field mode (the output)
via the quantum mechanical SQUID ring (figure~\ref{fig:schematic}(b)).
Here, as in  other work~\cite{r16}, we have modelled  this output mode
as a parallel $LC$ equivalent  circuit. However, in this paper we have
treated this circuit environment as classical since we wish to explore
a region  of the system parameter space,  involving large input-output
frequency ratios, which is of current experimental significance.  When
treated fully  quantum mechanically,  such simulations are  beyond the
reach of the computing power  available to us. As the essential result
of  this  paper,  we   show  that  the  non-perturbative  (non-linear)
interaction  generated by  the  SQUID ring  is  made manifest  through
energy conversion between  the input and output, even  if these differ
greatly  in  frequency.  In  part  this  has  been  inspired  by  past
experimental results  where we have  demonstrated~\cite{r17} that very
high  ratio frequency down  conversion can  occur between  a microwave
source  and a radio  frequency $\left(  \approx20\mathrm{MHz}\right) $
tank circuit  via the intermediary  of a very small  capacitance SQUID
ring~\cite{r17}.  In these previous  results we recorded  a frequency
down  conversion ratio of  500:1. In  recent, as  yet to  be published
data,  this  has  been  improved  upon by  refinements  in  electronic
technique to  yield a  ratio as high  as 6,500:1  between a a  few GHz
microwave  input field  and  a sub-MHz  tank  circuit. We  know of  no
classical  non-linear circuit  device that  could generate  such large
ratios  \cite{laser} and,  given the  very small  capacitance (niobium
point contact)  SQUID rings we were  using in our  experiments, it was
clearly of interest  to see whether a quantum  regime SQUID ring could
lead  to   this  massive  frequency  down  conversion.   In  a  recent
publication~\cite{r18} we  considered this  problem in the  context of
the persistent  current qubit circuit  model introduced by  Orlando et
al~\cite{r19},  where this  quantum element  acted as  the  medium for
coupling two classical  field environments together. Specifically, the
qubit was driven by a classical electromagnetic (em) field (the input)
which was used  to pump the qubit into an excited  state. To deal with
energy conversion  to an dissipative  output em mode, a  quantum jumps
approach was adopted to model the  decay of the excited qubit with the
energy dumped  into this  mode.  Using this  approach we were  able to
demonstrate small  ratio (500MHz to 300MHz)  frequency down conversion
from input  to output  through the intermediary  of the  quantum qubit
loop. In this work, with a  single weak link SQUID ring as the quantum
intermediary, we adopt a less  complicated model, where the SQUID ring
simply follows  Schr\"{o}dinger evolution without  the introduction of
quantum jumps.  We then show, to the limits of the computational power
available  to  us, that  a  two  orders  of magnitude  frequency  down
conversion (from microwave to radio frequencies) is possible. In doing
so  we  also  demonstrate  that  the  interaction  between  a  quantum
mechanical SQUID ring  and its classical circuit environment  is in no
way   trivial   and  requires   a   detailed   understanding  of   the
non-perturbative  properties of  SQUID rings.  We emphasise  that with
this  result  established,  and  without any  other  obvious  physical
constraints,  it  seems  reasonable   to  assume  from  a  theoretical
viewpoint that  even higher ratio  frequency down conversion  could be
realised  if  the necessary  computational  power  were available.  At
present,  apart from  essentially fixed  frequency masers,  sources of
quantum mechanical em fields  at microwave frequencies are not readily
available. Thus, given the  non-perturbative properties of SQUID rings
in the quantum  regime, it also seems reasonable  to consider how such
devices could  make manifest  their quantum mechanical  nature through
the interaction with classical em fields and field oscillator modes.

\begin{figure}
[ptb]
\begin{center}
\resizebox*{0.48\textwidth}{!}{\includegraphics{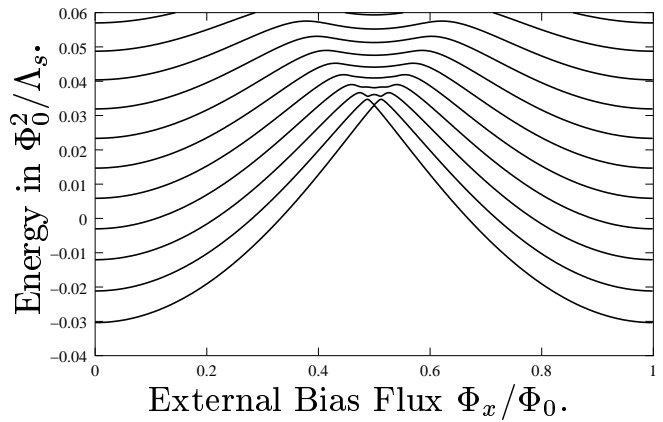}}
\caption{
  Plot of  the eigenenergies of  the SQUID ring Hamiltonian,  found by
  solving  the TISE,  as  a  function of  external  applied flux  with
  $\hbar\omega_{s}=0.006\Phi_{0}^{2}/\Lambda_{s}    $    and    $\hbar
  \nu=0.035\Phi_{0}^{2}/\Lambda_{s}$ \label{fig:eigenvalues}}
\end{center}
\end{figure}

In the well known lumped component model of a quantum mechanical SQUID
ring  (in this  work  one  Josephson weak  link  device, of  effective
capacitance  $C_{s} $,  enclosed by  a thick  superconducting  ring of
inductance $\Lambda_{s}$, with a classical magnetic flux of $\Phi_{x}$
applied  to  the   ring)  the  Hamiltonian  for  the   ring  is  given
by~\cite{r20,r21}
\begin{equation}
H_{s}=\frac{\hat{Q}_{s}^{2}}{2C_{s}}+\frac{\left(  \hat{\Phi}_{s}-\Phi
_{x}\right)  ^{2}}{2\Lambda_{s}}-\hbar\nu\cos\left(  \frac{2\pi\hat{\Phi}_{s}
}{\Phi_{0}}\right) \label{tiseham}
\end{equation}
where,  with a  circumflex denoting  operators,  $\hat{\Phi}_{s}$ (the
magnetic   flux  threading  the   ring  of   the  SQUID   device)  and
$\hat{Q}_{s}$  (the  Maxwell electric  displacement  flux between  the
electrodes  of  the  weak  link)  are  canonically  conjugate  quantum
variables such that $\left[ \hat{\Phi}_{s},\hat{Q}_{s}\right] =i\hbar$
with   $\hat{Q}_{s}\rightarrow-i\hbar  \partial/\partial\Phi_{s}$  and
$\Delta\hat{\Phi}_{s}\Delta  \hat{Q}_{s}\geq\hbar/2$. Here,  the third
term on the right hand side of (\ref{tiseham}) is due to the Josephson
phase  coherent  coupling energy,  with  matrix element  $\hbar\nu/2$,
arising  from  the  weak  link critical  current  $I_{c}=2e\nu$,  with
periodicity set  by the superconducting flux  quantum $\Phi_{0}=h/2e$. 
We assume that the ambient temperature $\left( T\right) $ of the SQUID
is  such that  $\hbar\omega_{s}\gg k_{B}T$  for a  characteristic ring
oscillator         frequency         of         $\omega_{s}/2\pi\left(
  =1/2\pi\sqrt{\Lambda_{s}C_{s}}\right)  $. Operating in  this quantum
regime the  time independent  Schr\"{o}dinger equation (TISE)  for the
SQUID ring then reads
\begin{equation}
\left[  \frac{\hat{Q}_{s}^{2}}{2C}+\frac{\left(  \hat{\Phi}_{s}-\Phi
_{x}\right)  ^{2}}{2\Lambda}-\hbar\nu\cos\left(  \frac{2\pi\hat{\Phi}_{s}
}{\Phi_{0}}\right)  \right]  \left\vert \psi_{\kappa}\right\rangle
=E_{\kappa}\left\vert \psi_{\kappa}\right\rangle \label{tiseeq}
\end{equation}
for   ring   eigenfunctions   $\left\vert   \psi_{\kappa}\right\rangle
$($\kappa=0$  the ground  state, $\kappa=1$  the first  excited state,
etc.)   and    ring   eigenenergies   $E_{\kappa}$    (as   shown   in
figure~\ref{fig:eigenvalues}),      these      eigenenergies     being
$\Phi_{0}$-periodic in the external magnetic flux applied to the ring.

\section{Dynamics of a Coupled Quantum SQUID Ring-Classical Resonator System}

\subsection{Adiabatic Regime}

In the time independent (adiabatic) case the experimentally accessible
state  is the  ground  state $\left(  \kappa=0\right)$  for which  the
expectation  value of  the screening  supercurrent flowing  around the
ring      is      $\langle      \hat{I}_{s}\left(      \Phi_{x}\right)
\rangle=-\left\langle  \partial\hat  {H}/\partial\Phi_{x}\right\rangle
=\left\langle           \left(          \hat{\Phi}_{s}-\Phi_{x}\right)
  /\Lambda_{s}\right\rangle$~\cite{r17,r20}.   For   relatively  large
values of  $\hbar\nu$ (and hence $I_{c}$) this  ground state screening
current   takes  the   form   of  a   rounded   sawtooth  centred   on
$\Phi_{x}=\Phi_{0}/2$  (modulo $n\Phi_{0}$,  $n$  integer). Thus,  the
ring  screening current  is clearly  a highly  non-linear  function of
$\Phi_{x}$,      as      is      its      magnetic      susceptibility
$\chi_{s}=\partial\langle\hat{I}_{s}\left(   \Phi_{x}\right)   \rangle
/\partial\Phi_{x}$.  At $\Phi_{x}=\left(  n+1/2\right)  \Phi_{0}$, $n$
integer,  the  ring  is   in  an  quantum  superposition,  with  equal
coefficients,  of  clockwise  and  anti-clockwise  screening  currents
(equivalently flux  states). A rounded sawtooth jump  in the screening
current, centred at half  integer bias flux, correspondingly generates
a narrow  positive spike in $\chi_{s}\left( \Phi_{x}\right)  $ at this
flux value.  At integer bias flux  $\left( \Phi_{x}=n\Phi_{0}\right) $
the susceptibility is a minimum, increasing monotonically to a maximum
at $\Phi_{x}=\left( n+1/2\right) \Phi_{0}$.  As we have shown, in this
simple ground state model of a quantum regime SQUID ring the evolution
of  its quantum  state  can be  monitored  through its  effect on  the
classical   dynamics    of   a   measurement    circuit   coupled   to
it~\cite{r15,r17,r20}.  Typically  this  takes   the  form  of  a  low
frequency, parallel $LC$ tank circuit inductively coupled to the SQUID
ring and  excited at,  or very close  to, its resonant  frequency. The
resonant frequency of this tank circuit $\left( \omega_{t}/2\pi\right)
$  is  usually extremely  low  compared  with  the natural  oscillator
frequency  of  the  SQUID  ring  (with the  implicit  assumption  that
$\hbar\omega_{t}\ll  k_{B}T$). This configuration  is very  well known
and  forms the basis  for one  type of  ultra sensitive  magnetic flux
detector    -   the    ac    or   radio    frequency   biased    SQUID
magnetometer~\cite{r21}. With  this tank  circuit playing the  role of
the  classical  measurement system  for  the  SQUID  ring, a  changing
$\chi_{s}\left(  \Phi_{x}\right)$ leads,  in the  ground state  of the
ring,  to  downward  shifts   (due  to  the  positive  $\chi_{s}\left(
  \Phi_{x}\right) $)  of the resonant  frequency of the tank  circuit. 
For  this  ground  state,  therefore,  following  the  change  in  the
frequency (and amplitude)  of the tank circuit resonance  allows us to
extract some information concerning the evolution of the quantum state
of  the   ring  (i.e.  the  coefficients  in   its  screening  current
superposition) as a function of $\Phi_{x}$.

\subsection{The Born-Oppenheimer Approximation}

\begin{figure}[!t]
\begin{center}
\resizebox*{0.48\textwidth}{!}{\includegraphics{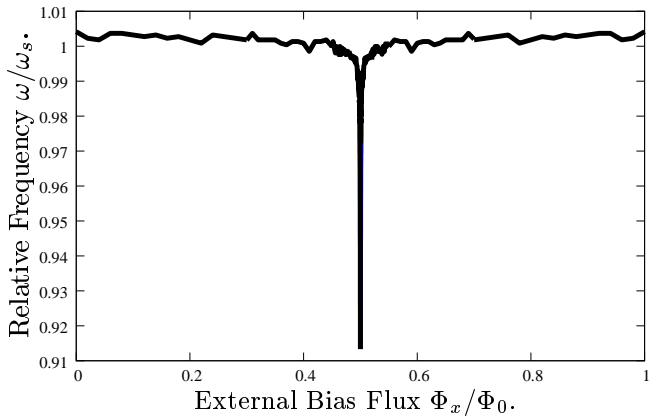}}
\caption{
  Frequency shift of  a 130MHz tank circuit ($L_{tc}=3\times10^{-8}H$,
  $C_{tc}=5\times10^{-11}F$), coupled  to the SQUID ring  of figure 2,
  as a  function of external applied  flux for the ring  in its ground
  state only  (figure 2(a)) calculated using  the the Born-Oppenheimer
  (adiabatic) approach.
\label{fig:freqShiftBO}}
\end{center}
\end{figure}

Given that the  oscillator frequencies of both the  SQUID ring and the
(classical)  tank circuit  resonator  are functions  of their  circuit
capacitances, it is,  in this limit, often convenient  to introduce an
approach  well known  in  atomic physics.  These  capacitances can  be
considered as  measures of the  effective mass of each  circuit (SQUID
ring and tank circuit  resonator). Crucially, for quantum regime SQUID
rings these capacitances (effective masses) will be markedly different
in magnitude,  e.g.  for  the ring typically  $\approx$ a  few $\times
10^{-15}\mathrm{F}$ and  for the tank circuit $\approx$  a few $\times
10^{-11}\mathrm{F}$,  or  more. This  low  mass-high  mass (here,  low
capacitance-high capacitance) situation was  dealt with many years ago
by  Born and  Oppenheimer for  nuclear-electron motion.  Adopting this
approach we  compute from solutions of  (\ref{tiseeq}) the expectation
value  of the  screening supercurrent  in  the SQUID  ring (fast)  and
substitute  this as  a feedback  term into  the classical  equation of
motion  for  the  tank  circuit  (slow). Provided  that  the  resonant
frequency of the  tank circuit is low enough, so  that to an extremely
good approximation the quantum SQUID ring remains adiabatically in its
ground state, the equation of motion for the coupled ring-tank circuit
system then reads~\cite{r15,r20}

\begin{equation}
C_{t}\frac{\partial^{2}\Phi_{t}}{\partial t^{2}}+\frac{1}{R_{t}}\frac
{\partial\Phi_{t}}{\partial t}+\frac{1}{L_{t}}\Phi_{t}=I_{\mathrm{in}}\left(
t\right)  +\mu\left\langle\psi\left\vert \hat{I}_{s}\right\vert \psi \right\rangle
\label{boequmotion}
\end{equation}
where  $L_{t}$  and  $C_{t}$   are,  respectively,  the  tank  circuit
inductance  and capacitance, $\Phi_{t}$  is the  magnetic flux  in the
tank  circuit inductor  and  $R_{t}$  is the  resistance  of the  tank
circuit on resonance. In this Born-Oppenheimer approximation we assume
that the ring remains in  one of its instantaneous energy eigenstates,
i.e.  solutions of (\ref{tiseeq});  in reality  this means  the lowest
energy state  (the adiabatic  limit). As we  have demonstrated  in the
literature~\cite{r22},  this  approximation  holds  very well  if  the
frequency  of  the  drive  is  very small  indeed  compared  with  any
separations, in frequency, between  the SQUID ring energy levels. This
certainly appears  to be the  case for radio frequency  (rf) circuits,
typically used  to probe single weak  link SQUID rings, as  well as in
their application  in SQUID magnetometry~\cite{r23,r24}.  We note that
since the SQUID  ring is macroscopic in nature (as,  of course, is the
tank circuit)  there exists a significant back  reaction between these
two  circuits,  as evidenced  by  the $\mu  \left\langle\psi\left\vert
    \hat{I}_{s}\right\vert     \psi     \right\rangle$     term     in
(\ref{boequmotion}).  In  practise  this  means that  the  SQUID  ring
(through the cosine in its Hamiltonian) generates a non-linear dynamic
in the classical  circuit environment to which it  is coupled, in this
case  a tank  circuit. For  small tank  circuit drive  amplitudes this
manifests itself  as a  frequency shift in  the power spectrum  of the
tank circuit as shown in figure~\ref{fig:freqShiftBO}

\section{Time Dependent Schr\"{o}dinger Equation Description - Exchange
Regions}

Following on from the  time independent Schr\"{o}dinger description of
a  SQUID  ring  (\ref{tiseeq}),  the  time  dependent  Schr\"{o}dinger
equation (TDSE) for the ring takes the form

\begin{equation}
\left[  \frac{\hat{Q}_{s}^{2}}{2C_{s}}+\frac{\left(  \hat{\Phi}_{s}-\Phi_x \right)  ^{2}}{2\Lambda_{s}}
-\hbar\nu\cos\left(  \frac{2\pi\hat{\Phi}_{s}}{\Phi_{0}}\right)  \right]
\left\vert \psi  \right\rangle =i\hbar\frac{\partial\left\vert
\psi  \right\rangle }{\partial t}\label{tdseeq}
\end{equation}
where  $\Phi_x=\Phi_{x}^{\mathrm{stat}}+\Phi\left( t\right)  $  is now
comprises  of a  time  dependent magnetic  flux $\Phi\left(  t\right)$
together  with a  static component  $\Phi  _{x}^{\mathrm{stat}}$.  The
intrinsically extremely  non-perturbative quantum nature  of the SQUID
ring would indicate that the application of em fields low in frequency
compared  with  the frequency  difference  $\left(  \times h\right)  $
between  adjacent  energy  levels  in  the ring  should  still  induce
transitions  between its  quantum levels~\cite{r25}.   Thus  for level
differences  of a  few hundred  GHz (quite  typical of  quantum regime
SQUID  rings), we  expect em  fields at  microwave frequencies  (a few
\textrm{GHz)} to  induce transitions, provided the  field amplitude is
high  enough~\cite{r25}. As  our example  in this  work we  consider a
SQUID ring with circuit  parameters commensurate with operation in the
quantum      regime\cite{r2,r26}.      For     this      we     choose
$C_{s}=5\times10^{-15}\mathrm{F}$,
$\Lambda_{s}=3\times10^{-10}\mathrm{H}$                             and
$\hbar\nu=0.035\Phi_{0}^{2}/\Lambda_{s}$,  the latter yielding  a weak
link  critical current $I_{c}\left(  =2e\nu\right) $  of $2\mathrm{\mu
  A}$. Furthermore, for an  oxide insulator, tunnel junction weak link
(oxide  thickness $\approx$ 1nm,  oxide dielectric  constant $\approx$
10) in  the  SQUID  ring,  this capacitance  corresponds  to  junction
dimensions $0.25\mu m$ square,  well within the capabilities of modern
microfabrication technology. In turn, these dimensions imply a maximum
supercurrent density in the weak link of around $4\mathrm{kAcm}^{-2}$,
a perfectly reasonable value for current experiments on quantum regime
SQUID  rings.   The   SQUID  ring  circuit  parameters  $C_{s}=5\times
10^{-15}\mathrm{F}$  and  $\Lambda_{s}=3\times10^{-10}\mathrm{H}$ lead
to a ring  oscillator frequency $\omega_{s}/2\pi =130\mathrm{GHz}$. If
we assume that the (planar) SQUID ring is fabricated in niobium, which
is  often the  case,  then  this ring  oscillator  frequency is  small
compared with the superconducting  energy gap in this material $\left(
  \approx1000\mathrm{GHz}\right)   $  at   low   reduced  temperatures
$T/T_{c}$,  where  $T_{c}$  is the  superconducting  critical
temperature   ($9.2K$  for   niobium)  for   an   operating  (ambient)
temperature in experiment typically around $40$mK.

Using the ring circuit parameters given above we showed in
\begin{figure}
[ptb]
\begin{center}
\resizebox*{0.48\textwidth}{!}{\includegraphics{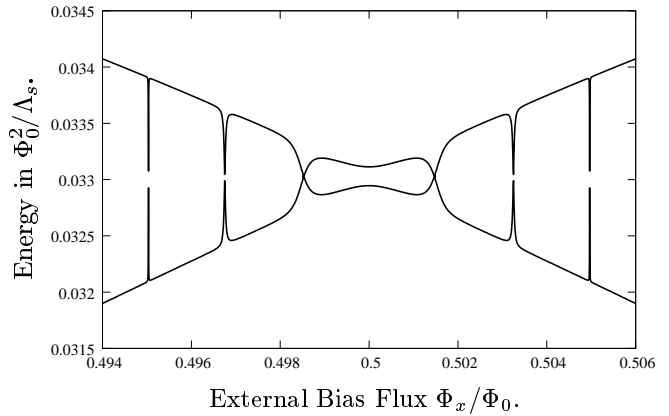}}
\caption{
  Time averaged energy expectation values  of the ring in the presence
  of  a microwave source  (input) at  a frequency  of 13GHz  using the
  first two energy eigenstates as initial conditions.\label{fig:tdse}}
\end{center}
\end{figure}
figure~\ref{fig:eigenvalues} the lowest  few energy eigenvalues of the
SQUID ring Hamiltonian, found by solving the TISE (\ref{tdseeq}), as a
function     of      the     static     applied      magnetic     flux
$\Phi_{x}^{\mathrm{stat}}$. Although such solutions of the TISE may be
adequate for  modelling this system  if the em fields  frequencies are
extremely small  compared with  frequencies $\left( \times  h\right) $
separating  the  ring  energy  levels (i.e.  in  the  Born-Oppenheimer
approximation, above),  in general we  must solve the TDSE  where time
dependent  fields  are  involved.   This is  clearly  demonstrated  in
figure~\ref{fig:tdse}.  Here,   using  the  the   ring  parameters  of
figure~\ref{fig:eigenvalues},  we  display  the time  averaged  energy
expectation  values  of  the  TDSE  (\ref{tdseeq}) as  a  function  of
$\Phi_{x}^{\mathrm{stat}}$,   for  an   applied  microwave   field  of
frequency $13$\textrm{GHz}, amplitude $0.001\Phi_{\emph{0}}$ and using
the first two eigenstates of the Hamiltonian as initial conditions. As
can  be seen,  for this  microwave  frequency and  amplitude there  is
significant energy exchange between these time averaged energies which
takes  place   over  a   a  very  narrow   range  of  the   bias  flux
$\Phi_{x}^{\mathrm{stat}}$.      We      term      these      exchange
regions~\cite{r16,r25}, or  equivalently, transition regions.  We note
that  at  the centres  of  these  exchange  regions the  corresponding
frequency   differences   between    the   energy   eigenenergies   in
figure~\ref{fig:eigenvalues}  are close  to integer  multiples  of the
applied microwave  frequency. At sufficiently high  em frequencies and
amplitudes several,  or many, exchange  regions are generated  with an
energy  spacing   between  adjacent  regions  very   close  to  $\hbar
\omega_{\mathrm{mw}}$,  where $\omega_{\mathrm{mw}}$ is  the (angular)
frequency  of  the   em  field.  As  we  have   shown  in  a  previous
publication~\cite{r16},  it  is in  these  exchange  regions that  the
interaction  between  a SQUID  ring  and one  or  more  em fields,  is
significant.  For example,  when a  field mode,  or modes,  is treated
quantum mechanically,  entanglement is generated between  the ring and
the  mode(s), reaching  (as with  the strength  of the  interaction) a
maximum in the centre of an exchange region. Given the result shown in
figure~\ref{fig:tdse},  where  it  is   clear  that  energy  is  being
exchanged between the SQUID ring  and the (classical) em field, we now
consider whether a similar exchange  can take place between this input
field and a classical output  field mode, coupled together through the
ring.

\subsection{Non-adiabatic Regime}

We  now investigate  a scenario,  accessible at  the current  level of
experimental  practise~\cite{r17}, of  a classical  field input  and a
classical  field  mode  output,  differing  widely  in  frequency.  We
demonstrate   that   energy   can    be   exchanged   via   a   highly
non-perturbative, quantum mechanical SQUID ring. In particular, and in
line  with this  experimental background~\cite{r17},  we  consider the
interaction and energy transfer via the SQUID ring between a microwave
input,  acting purely  as a  source, and  a lower  frequency, resonant
circuit output.  This models  the SQUID ring-tank  circuit (ac-biased)
magnetometer   configuration~\cite{r23}    in   a   non-adiabatic   em
(microwave)  field.  In  this  paper,   we  assume  that  there  is  a
macroscopically significant back reaction between the tank circuit and
the SQUID ring \cite{r15,r27}. However, for simplicity, we assume that
there is no direct coupling between the em input and output fields and
note that this is easy to realise experimentally.

In  order to  deal in  a general  manner with  this  coupled ring-tank
circuit  system we can  no longer  assume that  the behaviour  of this
system is  adiabatic in  nature. By implication,  this means  that the
tank circuit  resonant frequency  is sufficiently large  compared with
the frequency separations between the SQUID ring levels. Hence, we can
no longer invoke the  Born-Oppenheimer approximation. The TDSE for the
system then takes the form~\cite{r23}
\begin{widetext}
\begin{equation}
\left[  \frac{\hat{Q}_{s}^{2}}{2C_{s}}+\frac{\left(  \hat{\Phi}_{s}-\left(
\Phi_{x}^{\mathrm{stat}}+\Phi_{x}^{\mathrm{mw}}+\mu\Phi_{t}\right)  \right)
^{2}}{2\Lambda_{s}}-\hbar\nu\cos\left(  \frac{2\pi\hat{\Phi}_{s}}{\Phi_{0}}\right)  \right]  \left\vert \psi\left(  t\right)  \right\rangle
=i\hbar\frac{\partial\left\vert \psi\left(  t\right)  \right\rangle }{\partial
t}\label{pbotdseequ}
\end{equation}
\end{widetext}
where, as can be seen, the total external magnetic flux applied to the
SQUID  ring consists  of static  bias, microwave  excitation  and back
reaction       tank       circuit       contributions.      i.e.       
$\Phi_{x}=\Phi_{x}^{\mathrm{stat}}+\Phi_{x}^{\mathrm{mw}}+\mu\Phi_{t}$. 
Again, as with the Born-Oppenheimer approximation (\ref{boequmotion}),
we assume that this back reaction is macroscopically significant.

Given that the SQUID ring is now allowed to follow a
Schr\"{o}dinger evolution and retain its time dependence (\ref{boequmotion})
in this non-adiabatic regime, the equation of motion for the tank
circuit now becomes
\begin{equation}
C_{t}\frac{\partial^{2}\Phi_{t}}{\partial t^{2}}+\frac{1}{R_{t}}\frac
{\partial\Phi_{t}}{\partial t}+\frac{1}{L_{t}}\Phi_{t}=I_{\mathrm{in}}\left(
t\right)  +\mu \left\langle\psi\left(  t\right)  \left\vert \hat{I}_{s}\left(
t\right)  \right\vert \psi\left(  t\right)  \right\rangle\label{pboequmotion}
\end{equation}
where
\[
I_{s}=-\frac{\partial\hat{H}_{s}}{\partial\Phi_{x}}=\frac{\hat{\Phi}_{s}
-\Phi_{x}\left(  t\right)  }{\Lambda_{s}}
\]
We   solve    the   simultaneous   coupled    differential   equations
(\ref{pbotdseequ}) and (\ref{pboequmotion}), where the frequency ratio
between input microwave and output tank circuit modes is large $\left(
  \geq100\right) $.  In our opinion  these results shed light  on both
the  general  problem  of  the description  of  the  quantum-classical
interface and,  in particular,  the interaction of  non-linear devices
such as SQUID rings with their classical environments.

\section{Results}

\begin{figure}[!t]
\begin{center}
\resizebox*{0.48\textwidth}{!}{\includegraphics{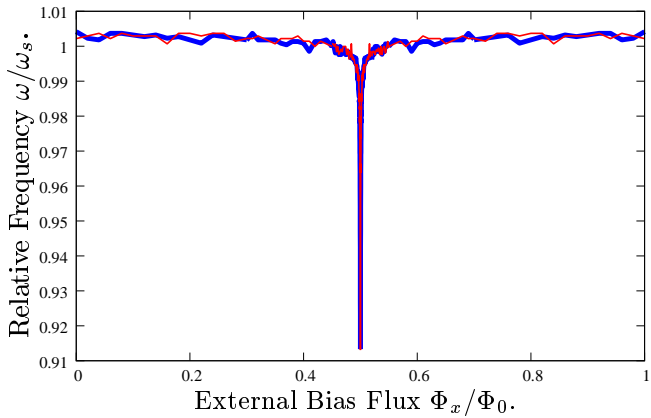}}
\caption{
  Frequency shift of  a 130MHz tank circuit ($L_{tc}=3\times10^{-8}H$,
  $C_{tc}=5\times10^{-11}F$), coupled  to the SQUID ring  of figure 2,
  as a  function of external applied  flux for the ring  in its ground
  state only (figure 2) calculated using the both the Born-Oppenheimer
  (in blue) and non-adiabatic (in red) approaches.
\label{fig:freqShift}}
\end{center}
\end{figure}

As a first check of the validity  of this model we compare it with the
established  Born-Oppenheimer approximation  for  the limiting  regime
where there are  no microwaves applied and the  tank circuit is driven
at  a   frequency  $f_{t}$   so  low  that,   to  an   extremely  good
approximation,  the SQUID  ring  remains adiabatically  in its  ground
state. This holds even  though, through solutions of (\ref{tiseeq}), a
whole  spectrum  of  ring   eigenstates  is  available.  In  order  to
demonstrate the  correspondence between these two models  in the limit
of  low   tank  circuit  drive  frequency,  we   calculate  the  $\Phi
_{x}^{\mathrm{stat}}$-dependent      frequency     shifts     $f\left(
  \Phi_{x}^{\mathrm{stat}}\right)  $  using  the  ring  parameters  of
figure~\ref{fig:eigenvalues}  and a  tank  circuit resonant  frequency
$130$\textrm{MHz} and a $\mu =0.01$. In figure~\ref{fig:freqShift} (as
we did in figure~\ref{fig:freqShiftBO}) we show the computed ring-tank
circuit resonant frequency shift in  this limit as a function of $\Phi
_{x}^{\mathrm{stat}}$ for both the Born-Oppenheimer approach (in blue)
and our generalised model (in red). It is quite apparent that there is
a very  high degree  of agreement between  the two approaches  in this
limit.
As our example of the frequency conversion process we consider a SQUID
ring  with  circuit  parameters   used  in  the  computed  results  of
figures~\ref{fig:eigenvalues}  and~\ref{fig:tdse}. For the  SQUID ring
parameters we have chosen the characteristic ring oscillator frequency
to be close to $130\mathrm{GHz}$. Referring to the computed results of
figure~\ref{fig:tdse}, we again choose the frequency for the microwave
source to be  $13\mathrm{GHz}$, i.e. an order of  magnitude lower than
the ring  characteristic frequency, and set the  tank circuit resonant
frequency   another   two   orders    of   magnitude   lower   (i.e.   
$130\mathrm{MHz}$).  Within  the  limits  of the  computational  power
available to us  this provides us with the  opportunity to demonstrate
extreme (1000:1), quantum SQUID ring mediated, energy down conversion.

With  the  results   of  figure~\ref{fig:freqShift}  in  support,  and
utilising   the   time   averaged   energy  expectation   results   of
figure~\ref{fig:tdse}  as a guide,  we then  proceeded to  compute the
energy transfer, via  the SQUID ring, from input  field to an undriven
output tank circuit  at selected flux bias points  close to and within
an exchange region in~\ref{fig:tdse}.
\begin{figure}[!t]
\begin{center}
\resizebox*{0.48\textwidth}{!}{\includegraphics{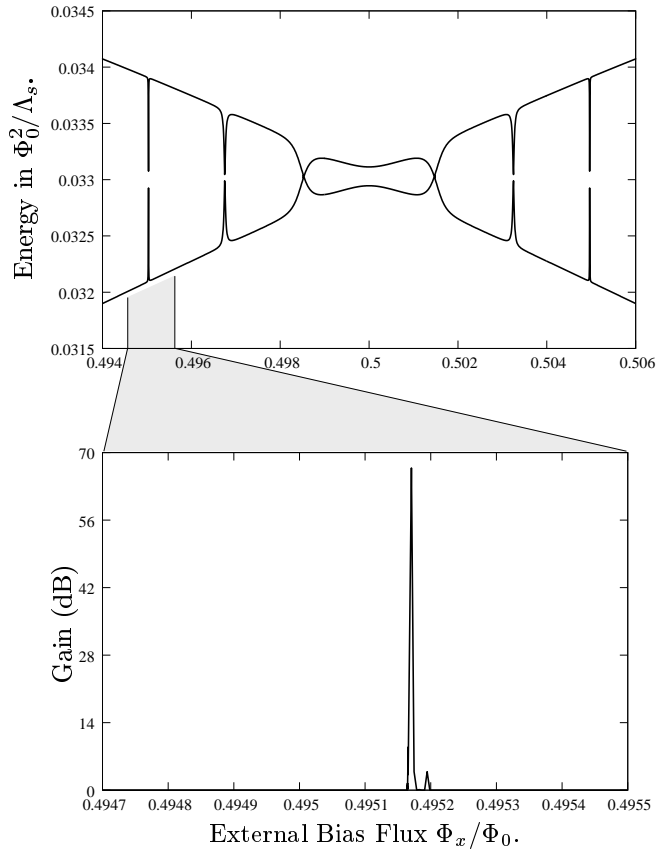}}
\caption{
  (top) reproduction of  figure~\ref{fig:tdse} illustrating the region
  of bias  flux used to  compute (bottom) the  power gain in  the tank
  circuit  (with respect  to  $\Phi_x^{\mathrm{stat}}=0\Phi_0$). Here,
  the microwave  source frequency was  13GHz which converted  into the
  output (tank circuit) oscillator mode of 130MHz. As can be seen, the
  peak power  at the  centre of the  exchange region  is $\approx70$dB
  above the background from outside the exchange region.}
\label{fig:powerGain}
\end{center}
\end{figure}
Our results are shown  in figure~\ref{fig:powerGain} for the region of
bias flux denoted using the time averaged energy expectation values at
the  top (which  reproduces figure~\ref{fig:tdse}).   As can  be seen,
outside the exchange region there  is no evidence of energy conversion
between  the  input microwave  field  and  the  output tank  circuit.  
However,  within  the  exchange   region  there  is  very  significant
conversion, reaching a maximum  gain ($\approx 70$dB above background)
at its centre.

The  results  we have  presented  in  this  paper demonstrate  that  a
input-output  frequency down  conversion of  a factor  of a  couple of
orders of magnitude can be  explained through the simple model we have
utilised. However, this does not constitute a theoretical upper limit,
only a  practical constraint arising  from the level  of computational
power available to  us. We also note that by symmetry,  and due to the
fact  that  the SQUID  ring  couples  different frequencies  together,
frequency  up  conversion  via  this  same mechanism  should  also  be
possible.

\section{Conclusions}

In this paper  we have demonstrated that the  interaction of a quantum
mechanical   SQUID  ring  with   classical  circuit   environments  is
non-trivial.  However,  in the  usual approach to  the influence  of a
classical (and dissipative) circuit  environment on the time evolution
of a quantum mechanical SQUID  ring~\cite{r12}, it is assumed that the
environment can be  modelled by a bath of  linear harmonic oscillators
linearly coupled to  the ring~\cite{CL83}. This need not  be the case. 
The highly  non-perturbative nature of  the SQUID ring in  the quantum
regime (and other  Josephson weak link based circuits)  means that the
ring-environment interaction  can be very  non-linear and may  lead to
unexpected  results. One  of these  is clearly  the  extreme frequency
ratio  conversion  possible between  classical  fields  via a  quantum
mechanical   SQUID   ring.   Since   the  problem   of   environmental
(dissipative,  decohering) effects  is  so central  to the  successful
implementation  of quantum technologies,  it is  our opinion  that for
non-linear devices such as SQUID rings the role of the environment has
to be reconsidered within this non-perturbative (non-linear) context.

On a  purely experimental level  the theoretical results  generated in
this  paper indicate that  quantum SQUID  rings can  be used  for very
large frequency  ratio down conversion between  classical fields. This
seems to be supported by  experiment~\cite{r17} and may prove to be of
considerable  practical  significance,  especially  with  the  current
interest in classical THz communications technologies~\cite{DLP04}.

\section{Acknowledgements}

We would like to thank the Engineering and Physical Sciences Research Council
for supporting the work presented in this paper through its Quantum Circuits
Network programme.

\end{document}